\documentclass[letter]{aa} % for the letters 
\usepackage{graphicx}
\usepackage{txfonts}
\usepackage{bm}
\usepackage[authoryear]{natbib}
\bibpunct{(}{)}{;}{a}{}{,}
\begin{document}
   \title{Zeeman-Tomography of the solar photosphere}

   \subtitle{3-Dimensional surface structures retrieved from Hinode observations}

   \author{T. A. Carroll
          \and
          M. Kopf
          }

   \offprints{T.A. Carroll}

   \institute{Astrophysikalisches Institut Potsdam, 
              An der Sternwarte 16, D-14482 Potsdam, Germany\\
              \email{tcarroll@aip.de}
             }

   \date{Received December 5, 2007; accepted December 21, 2007}

% \abstract{}{}{}{}{} 
% 5 {} token are mandatory
 
  \abstract
   {}
   {The thermodynamic and magnetic field structure of the solar photosphere is analyzed by means of a novel 
   3-dimensional spectropolarimetric inversion and reconstruction technique.}
   {On the basis of high-resolution, mixed-polarity magnetoconvection simulations, we used an artificial 
   neural network (ANN) model
   to approximate the nonlinear inverse mapping between synthesized Stokes spectra and the underlying 
   stratification of atmospheric parameters like temperature, line-of-sight (LOS) velocity and LOS magnetic field. 
   This approach not only allows us to incorporate more reliable physics into the inversion process,
   it also enables the inversion on an absolute geometrical height scale, which allows 
   the subsequent combination of individual line-of-sight stratifications to
   obtain a complete 3-dimensional reconstruction (tomography) of the observed area.}
   {The magnetoconvection simulation data, as well as the ANN inversion, have been properly processed to be 
   applicable to spectropolarimetric observations from the Hinode satellite.
   For the first time, we show 3-dimensional tomographic reconstructions 
   (temperature, LOS velocity, and LOS magnetic field) of a quiet sun region observed by Hinode.
   The reconstructed area covers a field of approximately $12\;000 \times 12\;000$ km and a height range of  
   $510$ km in the photosphere. 
   An enormous variety of small and large scale structures can be identified in the 3-D reconstructions. 
   The low-flux region ($B_{mag} = 20G$) we analyzed exhibits a number of \emph{tube-like} magnetic structures with 
   field strengths of several hundred Gauss. Most of these structures rapidly loose their strength with height and only 
   a few larger structures can retain a higher field strength to the upper layers of the photosphere.}
   {}


   \maketitle
%
%________________________________________________________________

\section{Introduction}
\label{Sect:1}
There has been a growing interest in the (small-scale) magnetic field from low-flux regions 
(i.e. internetwork) over the recent years 
\citep[e.g.][]{Lites02,Lites04,Socas04,Khom05,Dom06a,Lopez07},
not least because these regions make up most of the solar surface and may therefore carry
most of the unsigned magnetic flux and magnetic energy \citep{Sanchez04}.
Spectropolarimetric observations and their analysis are extremely challenging in these low-flux regions
because of the weakness of the polarimetric signals and because of the complex arrangement
and small filling factor of the underlying magnetic field.
Despite these complexities, the analysis of spectropolarimetric observations in terms
of a Stokes profile inversion may still be the most promising way
to reveal the magnetic properties of these low-flux regions.
A number of different inversions have recently been applied to spectropolarimetric observations from 
internetwork regions
\citep[e.g.][]{Sanchez00,Socas02,Dom06b,Orozco07b} and could retrieved one-dimensional 
(depth independent) estimates of the magnetic field strength for each line-of-sight (LOS). 
But, as model assumptions and constraints are inevitable, the question
arises to what extent these assumptions are physically meaningful and how inversions can yield 
reliable results. In this context, note the potential degeneracy that can be caused by the subtle 
interplay between thermodynamics and magnetic fields \citep{Gonzales06}.

To implement more physical realism and more reliable contraints into the inversion process we have 
incorporated the most recent 
magnetohydrodynamic (MHD) simulations \citep{Vogler05} into an artificial neural network (ANN) inversion
\citep{Carroll01b}. This inversion which takes into account the full dynamical
situation present in the MHD data, allows us to extract the depth stratification of the temperature, 
LOS velocity, and the LOS magnetic field. Moreover, this approach facilitates the inversion on an absolute
height scale, and thus the 3-dimensional reconstruction (tomography) of the entire field-of-view (FOV).

\section{Basic strategy}
\label{Sect:2}
Our approach is based on the assumption that MHD simulations have reached a degree of realism
that allows them to provide a meaningful and quantitative representation of magnetoconvective processes 
in the solar photosphere.
If this assumption is approximately true, MHD simulations can be used to generate 
a large and statistically significant sample of possible magnetoconvective scenarios
such that a machine learning algorithm \citep{Bishop95} can
extract the underlying generator (function), which describes the relationship between the 
Stokes spectra and the atmospheric parameters. 

\section{MHD-Simulations, spectral synthesis, image and data degradation}
\label{Sect:3}
To provide a large and quasi-realistic sample of possible magnetic and thermodynamic scenarios for the 
training process of the ANNs, we have used 
recent high-resolution, non-grey, mixed-polarity MHD simulations \citep{Vogler05}, 
which comprises three snapshots of different magnetic regimes. One snapshot has an average unsigned magnetic field strength
of $\langle B \rangle$ = 22 G, and represents a low-flux region typical for the internetwork; a second one
covers a medium range with an average unsigned field strength of $\langle B \rangle$ = 50 G; and a third one 
represents a strong-flux scenario with $\langle B \rangle$ = 144 G, comparable to an active plage region.
Each simulation snapshot covers a computational domain of 6000 $\times$ 6000 $\times$ 1400 km$^3$, which is 
discretized by 576 $\times$ 576 $\times$ 100 grid points.
Based on the depth stratification of the individual MHD snapshots we have carried out the
synthesis of the Stokes spectra for each LOS (assuming disc center observations) for the  
two iron lines \ion{Fe}{i} 6301.5 \AA\ and  \ion{Fe}{i} 6302.5 \AA\ .

To make synthetic profile images comparable to observations from the Solar Optical Telecope (SOT) 
and the spectropolarimeter (SP) of Hinode \citep{Lites01,Shimizu04}, 
we followed the same procedure as described by \citet{Orozco07a}. 
We first degraded the synthetic profile images by an appropriate modulation transfer function (MTF) 
which also accounts for the effect of the finite CCD pixel size. We then resampled (rebinned) the 
profile images according to the CCD pixel size of 0.16\arcsec $\times$ 0.16\arcsec and finally convolved the 
individual profiles with a Gaussian function of 30 m\AA\ FWHM to account for the limited resolving power of the 
spectrograph.

To assure the agreement between the individual Stokes spectra and the underlying atmospheric parameters on
a spatial scale of 0.16\arcsec $\times$ 0.16\arcsec, we applied to the MHD snapshots,
which contain the atmospheric data (temperature, velocity and magnetic field)
the same resampling and averaging (rebinning) procedure as for the spectral images.
Thus, each of the 100 depth level in the original MHD snapshots 
has been resampled to obtain a \emph{mean} depth stratification (for each parameter) in each resampled pixel. 
Note, that the original sampling of the high-resolution  MHD simulation corresponds to a spatial scale
of 0.014\arcsec $\times$ 0.014\arcsec such that the resampled profile images and data layers 
comprises more than  11 $\times$ 11 pixel of the original MHD snapshot. 
This \emph{smearing} can already lead to a significant reduction in the peak values of the
magnetic field strength (up to 25\% in our case).
 
\section{Artificial neural network inversion and tomography}
\label{Sect:4}
Our inversion is based on multi-layer perceptrons (MLPs), a popular type of artificial neural networks 
\citep{Bishop95} that have already been successfully applied in the field of Stokes profile inversions 
\citep{Carroll01b,Socas05}.
Based on the data provided by the MHD simulations and the synthetic Stokes profiles,
we have trained three different MLPs to approximate the nonlinear mapping between the 
Stokes profiles of the two iron lines and the underlying stratification of the temperature, the LOS velocity, 
and the LOS component of magnetic field vector.
From each of the three available MHD snapshots we extracted a set of 5000 Stokes spectra together 
with the underlying depth stratification of the atmospheric data. The 5000 different positions were
obtained by randomly placing a sliding window (11 $\times 11$ pixels) on the original snapshots.
This procedure allowed us to compile a dataset of 15$\:$000 depth stratifications (for all 3 parameters) 
together with their corresponding (degraded) synthetic Stokes spectra. 
To each spectra we added Gaussian noise of a level of 10$^{-3}$, which is comparable to the expected 
noise level of Hinode/SP \citep{Lites01}.
Since we are, in particular, interested in analyzing Stokes profiles from low-flux regions we concentrate
on Stokes $I$ and Stokes $V$ profiles only and refrain from analyzing
the weak linear polarization signal in this work.

Since the intrinsic dimensionality of the Stokes profiles is usually much less than the typical wavelength
sampling \citep{Asensio07}, we have applied a principal component analysis (PCA) \citep{Rees00} to our 
database of synthetic Stokes profiles. 
The PCA allows us to describe efficiently each of the two iron Stokes $I$ profiles 
by using the first 10 eigenvectors (principal components) only. This provides a RMS reconstruction error 
as low as 10$^{-4}$ for the entire database.
Due to the higher relative noise level we used only 5 principal components for each of the Stokes $V$ profiles, which 
still allows for a RMS reconstruction error smaller than 5 $\times$ 10$^{-4}$,  and thus 
well below the assumed noise level.

Even after the resampling and averaging of the atmospheric parameters, the individual stratifications in 
each pixel still exhibit a complex and sometimes irregular variation along the LOS. To efficiently 
but also accurately describe the individual depth stratifications we have to 
choose an appropriate and flexible description. For this reason, we took advantage of the exhaustive statistics
provided by the MHD simulations (15$\:$000 stratifications) to apply a 
principal component analysis (PCA) to the depth stratification
of the atmospheric parameters. We have decomposed the stratifications of the 
temperature, the LOS velocity and the LOS magnetic field for
the upper 510 km, which comprises 37 grid point in the original MHD snapshots.
After calculating the PCA for the entire database each atmospheric stratification could be sufficiently 
described by using only the five principal components (eigenvectors).
The appropriateness of this parameterization can be 
evaluated by reproducing the 37 original grid points for each LOS from the five projection coefficients.
The mean reproduction error per grid point for all available 15$\:$000 depth stratifications are 
1.05 Gauss for the LOS component of the magnetic field vector,
59.43 m/s for the LOS velocity, and 37.84 K for the temperature.

Three separate MLPs are used to retrieve the temperature, velocity, and magnetic field information along
the LOS.
One MLP is trained to approximate the mapping between the two PCA decomposed Stokes $I$ profiles of the iron lines
and the PCA decomposed temperature stratification;
another one for the 
mapping between the decomposed Stokes $I$ profiles and the decomposed velocity stratification; and a 
third one for the mapping between the two decomposed Stokes $V$ profiles and the decomposed magnetic field
stratification.
The training process of the MLPs follows that described in detail by \citet{Carroll01a, Carroll01b}.
Since the original atmospheric stratifications in the MHD simulations are given on an absolute geometrical height scale, 
the MLPs describe the inverse mapping on the same absolute scale that comprises a height range of 510 km and 
corresponds to a mean logarithmic optical depth range between 0.6 and -4.0.
This fact facilitates the subsequent combination of the individual results from the 
LOS inversion to yield a complete 3-dimensional reconstruction of the observed area.

The trained MLPs have been tested by applying \emph{unknown} synthetic 
Stokes $I$ and $V$ profiles from a MHD snapshot ($\langle B \rangle$ = 22 G)
that were not part of the training database. 
The MHD test case comprises 50 $\times$ 50 $\times$ 37 (92$\;$500) grid points for each of the three
atmospheric parameter.
Before the synthetic Stokes spectra of this test case were presented to the MLPs the profiles went
through the same degradation and decomposition process as the training data and were 
additionally corrupted by noise of 10$^{-3}$. 
The individual MLPs then completely reconstructed the atmospheric structure of the MHD test case
from the Stokes $I$ and $V$ spectra of the two iron lines. 
The mean absolute error per grid point (with respect to the MHD test snapshot) 
for the test case was 73.51 K for the temperature, 195.08 m/s for the velocity,
and only 5.79 G for the LOS magnetic field component. 
A detailed analysis of the individual MLPs and the presentation of the impressive results 
will be given in a forthcoming paper \citep{Carroll08}. 

\section{The real case, tomography of a quiet Sun region}
\label{Sect:5}
\begin{figure}
\centering
\includegraphics[width=4.45cm,height=4.45cm]{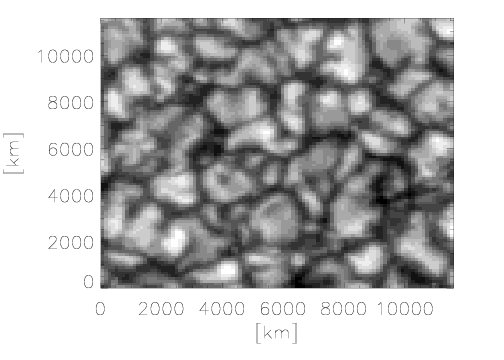} 
\includegraphics[width=4.45cm,height=4.45cm]{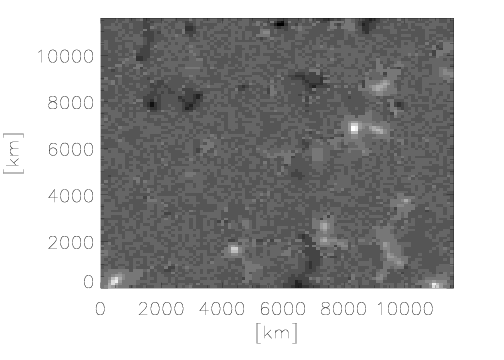}
\thicklines
\put(-4,37){\vector(-1,-1){10}}
\put(-17,25){\circle{7}}
\caption{Continuum intensity (left) and magnetic flux density (right) for the analyzed quiet-sun region.
The small circle on the lower right in the flux density figure indicates a strong flux structure which
will be shown in greater detail in the following tomographic reconstruction}
\label{Fig:1}
\end{figure}

\begin{figure}[t]
\centering
\includegraphics[width=10cm,height=8cm]{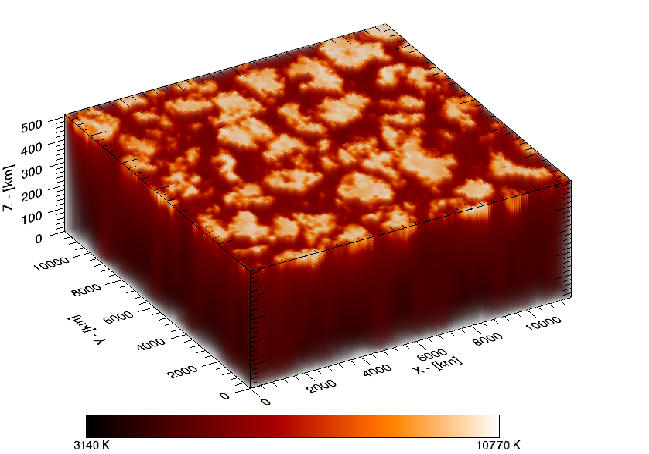} 
\caption{Tomographic reconstruction of the temperature for the analyzed photospheric region. For the purpose
of a better visualization, the reconstructed 3-dimensional surface volume is shown in bottom view.
Note how the bright (hot) and the dark (cool) areas in the deep layers (here on top)
nicely reproduce the granulation pattern (Fig.\ref{Fig:1}).}
\label{Fig:2}
\end{figure}

\begin{figure}
\centering
\includegraphics[width=10cm,height=8cm]{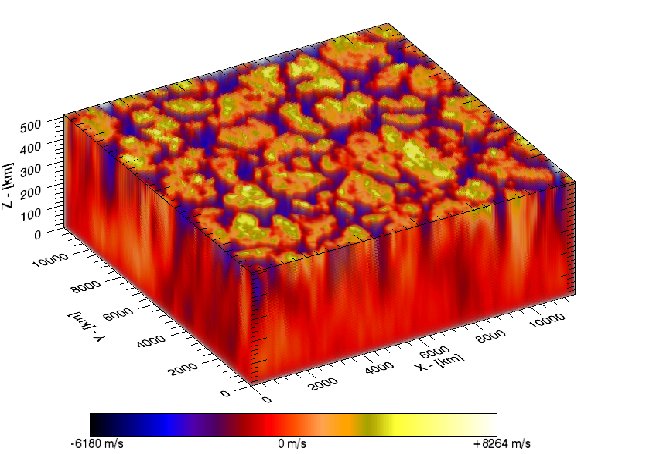} 
\caption{3-Dimensional reconstruction of the LOS velocity. As in Fig.\ref{Fig:2} the reconstruction shows the
deepest layer on top. Yellow and orange colors indicate strong upflows (here in the downward direction), and
blue and dark colors indicate downflows (here upward direction). Note the onset of strong downflows
(blue) in the mid-photosphere}
\label{Fig:3}
\end{figure}

\begin{figure}
\centering
%[scale=0.5]
\includegraphics[width=10cm,height=8.5cm]{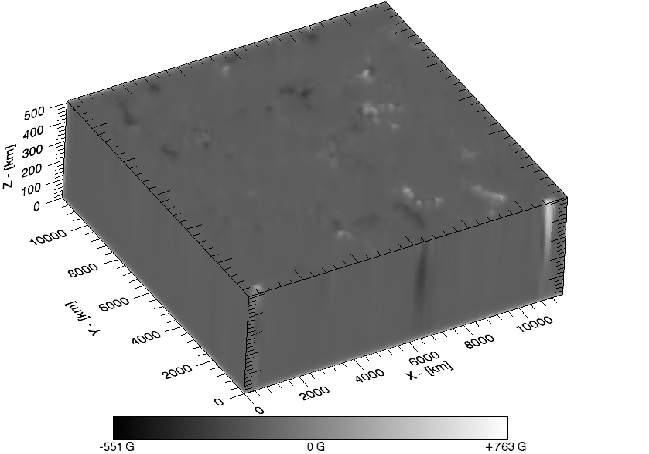} 
%\thicklines
\put(-46,111){\oval(45,65)}
\caption{The 3-dimensional reconstruction of the LOS magnetic field. As in Fig.2 \& Fig.3 the 
reconstruction is shown in a bottom up view. Bright and dark colors represent the 
different magnetic polarities. The marked region which comprises the magnetic structure
encircled in Fig.\ref{Fig:1} provides a cross-section view of this structure.}
\label{Fig:4}
\end{figure}

\begin{figure}
\centering
\includegraphics[width=5cm,height=5cm]{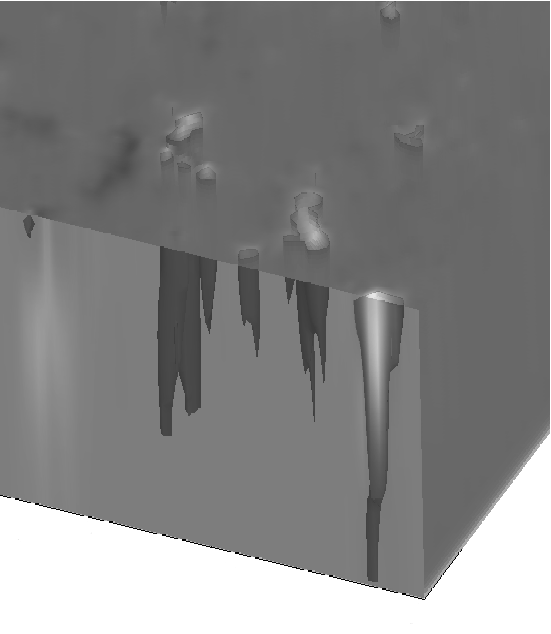} 
%\thicklines
\put(-44,76){\circle{14}}
\caption{A close up view of the magnetic structure, marked in Fig.\ref{Fig:4}, shown with transparent X-Z plane 
and iso-surfaces of 175 G. It shows a cross-section through the tube-like magnetic structure.
The highest field strength of 763 G is reached within the encircled region.}
\label{Fig:5}
\end{figure}

\begin{figure}
\centering
\includegraphics[width=5cm,height=5cm]{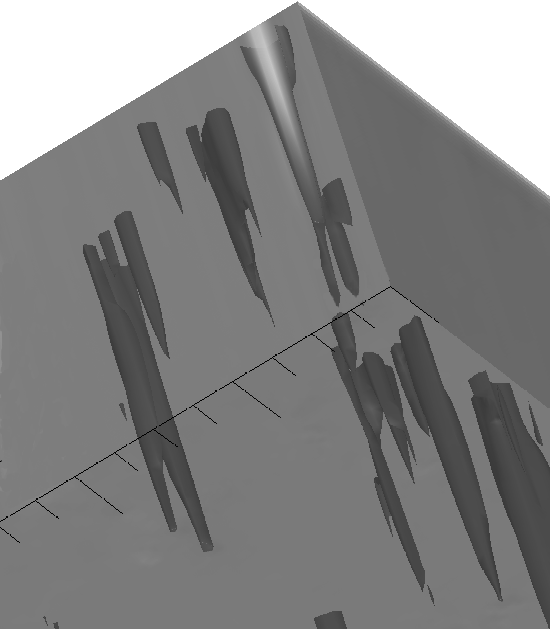}
\caption{Another close up view of the same structure, but from a different viewing angle.
The data volume is rotated around the X and Y axes (lifted)
to allow a view through the transparent bottom plane of the box. This view corresponds 
to an oblique view from the \emph{top} of the atmosphere.}
\label{Fig:6}
\end{figure}

The observation we used were taken on March 10, 2007 with the spectropolarimeter (SP)
aboard Hinode. The scan comprises a quiet sun region at disc center and covers a FOV of 302" $\times$ 162".
The SP recorded the Stokes spectra of the iron line pair at FeI 6301 \AA\ with a wavelength sampling of 2.15 pm.
The data have been corrected for dark current, flat field, and instrumental effects as
described by \citet{Lites07}. We conducted a wavelength calibration and continuum normalization
according to the procedure described by \citep{Jurcak07}.
As explained in the preceding section we designed the preparation of the MHD data and spectra as well
as the neural network training to match Hinode/SP observations. Thus, we
could directly decompose the measured Stokes profiles and apply them to the individual MLPs.
Note that due to the processing of the MHD data (resampling) the ANN inversion 
operates on a spatial scale that corresponds to the CCD pixel size of 
0.16'' $\times$ 0.16'' and therefore facilitates a one-component inversion on that scale.
The inversions and 3-dimensional reconstructions were made for the entire FOV but for the sake of clarity 
and a better illustration of the results we confine ourselves to a subregions
that covers an area of 100 $\times$ 100 pixel (16" $\times$ 16" $\approx$ 11$\:$600 $\times$ 11$\:$600 km).
The continuum and the magnetic flux density estimation (as retrieved from the 
measured Stokes $V$ profiles) are shown in Fig.\ref{Fig:1}.
From the magnetic flux density estimation we deduce a mean unsigned flux density 
$B_{mag}$ of approximately 20 G for the subregion. 
%[width=4.45cm,height=7.0cm]
The results of the temperature and the LOS velocity tomography are shown in Fig.\ref{Fig:2} and Fig.\ref{Fig:3}.
For the purpose of showing the rich and detailed structure of the 3-dimensional reconstructions, the data volumes 
are shown in a bottom view (upside down) such that the deepest layers of the photosphere ($<$log$(\tau)$$>$ $\approx$ 0.6)
are on top.
The tomography of the LOS magnetic field for the subregion is shown in Fig.\ref{Fig:4}. This
reconstruction is also in a bottom up view to give a better illustration of the many details 
in the lower atmosphere. The bipolar features of the flux density estimation
of Fig.\ref{Fig:1} are well reproduced, but in much greater detail
and, moreover, many fine scale structures are visible at the bottom layer at $<$log$(\tau)$$>$ $\approx$ 0.6.
The whole 3-dimensional reconstruction comprises 370$\:$000 (100 $\times$ 100 $\times$ 37) grid points.
In the front X-Z plane of Fig.\ref{Fig:4}, we see that our tomography reveals the slice view of three magnetic structures.
A close-up view of one of these magnetic structures
is shown in Fig.\ref{Fig:5} and Fig.\ref{Fig:6}. 
The position of this structure is marked by the encircled region in the flux density image (Fig.\ref{Fig:1})
and in the 3-D reconstruction (Fig.\ref{Fig:4}). 
The side wall (X-Z plane) of Fig.\ref{Fig:5} as well as the bottom plane (X-Y plane) in Fig.\ref{Fig:6}, are 
made transparent to allow a view into the \emph{inner} part of the reconstructed data volume.
Isosurfaces of +175 G are made visible, revealing the existence of many tube-like magnetic structures. 
One can see that the
depicted strong magnetic feature in Fig.\ref{Fig:5} and Fig.\ref{Fig:6} exhibits not only a tube-like character
but also an \emph{internal} structure with vertical and lateral gradients (increased whitening towards the inner). 
This internal structuring is also characterized by the increased narrowing of the flux structures
toward the upper atmosphere (again note, the bottom up view of the representation).
It is also conspicuous that many of the isosurfaces do not 
reach the upper layers of the atmosphere, indicating the rapid decrease of magnetic field strength with height.
This fact will be highlighted in our upcomming paper \citep{Carroll08} in which we
present the first 3-dimensional empirical probability density functions (pdfs) for 
the magnetic field strength. These pdfs reveal the characteristic height dependent changes 
(shift of mean and varaiance) of the magnetic field strength.
In this context, it is also interesting to note that the mean unsigned flux at the bottom layer in the 
reconstructed volume decreases from 15.88 G to 4.24 G at the top layer (510 km above).
This also indicates that much of the magnetic flux does not reach the upper photosphere in the form of
magnetically dense flux structures.
The peak field strength of 763 G in the analyzed 3-dimensional region is encountered 
on the footpoint of the depicted magnetic structure and is marked by the circle in Fig.\ref{Fig:5}.
Note, the maximum value in the flux density estimation (Fig.\ref{Fig:1}) 
only yields 297 G for this structure. 

\section{Discussion and conclusion}
\label{Sect:6}
We have presented an inversion approach based on artificial neural networks that incorporates the knowledge from
high-resolution, mixed-polarity magnetoconvection simulations. This approach does not only allow the 
accurate determination of the depth stratification of various atmospheric parameters, it also facilitates the 
complete 3-dimensional reconstruction of the analyzed atmosphere. The tomography retrieves the information from 
a surface layer of 510 km in the photosphere, which covers a mean logarithmic optical depth range from 
0.6 to -4.0. The presented inversion approach allows a unique 3-dimensional investigation of 
individual photospheric structures.
Based on Hinode/SOT observations of a quiet sun region, this work has demonstrated the feasibility 
of a tomographic reconstruction for the temperature, LOS velocity, and magnetic field.
Nevertheless, it is important to emphasize that this approach is based on the premise that
MHD simulations have reached a sufficient degree of realism and that the simulation snapshots used in this
work provide a statistically relevant sample of possible physical scenarios encountered in the solar photosphere. 
If these conditions are met, the presented ANN approach and tomography provides a powerful 
and exciting new diagnostic for the magnetic field of the solar photosphere. 

\begin{acknowledgements}
      We thank M. Sch\"ussler for providing the snapshots of the MHD simulations and 
      J. A. Bonet and S. Vargas for providing the MTF and their code to appropriately 
      degrade our synthetic Stokes profiles. We also want to thank the referee A. Asensio Ramos 
      for many helpful and constructive comments.
      
\end{acknowledgements}

\end{document}